\documentclass[11pt]{article}
\usepackage[dvips]{graphicx}

\usepackage{epsfig}
\usepackage{graphicx}
\usepackage{indentfirst}
\usepackage{amsmath}
\usepackage{amsfonts}
\usepackage{amssymb}

\begin{document}

\begin{center}
\begin{flushright}\begin{small}    UFES 2011
\end{small} \end{flushright} \vspace{1.5cm}
\huge{New Black Holes Solutions in a Modified Gravity} 
\end{center}

\begin{center}
{\small   M. Hamani Daouda $^{(a)}$}\footnote{E-mail address:
daoudah8@yahoo.fr}\ ,
{\small  Manuel E. Rodrigues $^{(a)}$}\footnote{E-mail
address: esialg@gmail.com}\ and
{\small   M. J. S. Houndjo $^{(b)}$}\footnote{E-mail address:
sthoundjo@yahoo.fr} \vskip 4mm

(a) \ Universidade Federal do Esp\'{\i}rito Santo \\
Centro de Ci\^{e}ncias
Exatas - Departamento de F\'{\i}sica\\
Av. Fernando Ferrari s/n - Campus de Goiabeiras\\ CEP29075-910 -
Vit\'{o}ria/ES, Brazil \\
(b) \ Instituto de F\'{i}sica, Universidade Federal da Bahia, Salvador, BA, Brazil\\
\vskip 2mm

\end{center}

\begin{abstract}
We present some basic concepts of a theory of modified gravity, inspired by the gauge theories, where the commutator algebra of covariant  derivative gives us an added term with respect to the General Relativity, which represents the interaction of gravity with a substratum. New spherically symmetric solutions of this theory are obtained and can be view as  solutions that reproduce, the mass, the charge, the cosmological constant and the Rindler acceleration, without coupling with the matter content, i.e., in the vacuum. 
\end{abstract}

Pacs numbers: 04.50.Kd, 04.70.Bw, 


\section{Introduction}
Through the need for modeling the phenomena and symmetries that the nature can present, the mathematical tools more used were the group. A group obeys to some specified algebraic properties that characterize it. Several theories of group that served to model the symmetries of various quantum phenomena are well known and are for great success \cite{nakamura}. The standard model of interactions is basically divided into the  Electroweak Interactions model, developed by Abdus Salam, Sheldon Glashow and Steven Weinberg \cite{gautam}, whose symmetries group is  $SU(2)\times U(1)$, and the Strong Interactions characterized by the symmetries group $SU(3)$ \cite{nakamura}. Note that this model does not include the gravitational interaction.\par
Various models of grand unification emerged in order to provide the unification of the elementary interactions. Some of the most important were those of  Weyl \cite{weyl}, Supergravity \cite{wess}, Strings and Superstrings \cite{ortin} and Loop Quantum Gravity \cite{lee}. In all these models the group theory has an important role; it is an ingredient of the description  of the known revealed symmetries of the nature.\par
As in the case of quantum or microscopic phenomena, the comprehension of the content and the evolution of the universe has grown in recent years. With a great advance in measurements of the cosmological data, and the increase of cosmological and astrophysical measures, it has been obtained the acceleration and the strange content of the universe, which according to the General Relativity (GR), are the dark matter and the dark energy. A not so small class of modifications of the GR has emerged as an attempt of a new approach to explain consistently with more recent data, as  WMAP \cite{wmap} and type Ia Supernovae \cite{riess}. The theories of modified gravity most widely used as alternatives that allow us to understand the analyse of the expansion phenomenon of the universe or the dark matter content are the TEVES \cite{bekenstein}, the $f(R)$ \cite{capozziello,odintsov}, the $f(G)$ \cite{sergei} (with f(R,G) \cite{antoniadis} and f(R,T) \cite{sergei2}) and the f(T) \cite{stephane} ( generalization of the Teleparallel Theory \cite{pereira}). \par
Inspired by Lorentz-Yang-Mills gauge theory, whose symmetries group is $SO(3,1)$, Camenzind \cite{camenzind}, in different order, constructs a modification of the GR in which the Lagrangian contains a quadratic term of the scalar curvature and the quadratic contraction of the Ricci tensor. Using the Levi-Civita connection and an analogy with Yang-Mills theory, he obtains the field equations in which the energy-momentum tensor is not conserved. The conserved quantity in his theory, due to the Bianchi identity, is the 4-current associated with the matter content, i.e, the energy momentum tensor.\par  
Similarly, later, inspired by the foundations of the gauge theory, Resconi et al \cite{resconi1} apply a new methodology for the commutators algebra of covariant derivatives which is the construction of a modification of the GR, and which leads to the GR in a particular case. In the same way as Camenzind \cite{camenzind}, Resconi et al found an imposition on the theory, that is, the 4-current acts as the source of the matter content and is conserved, while the energy-momentum tensor is not. Through an approximate analysis, they get for the static and spherically symmetric case, an increase in  mass of a Schwarzchild  type black hole, which is an correction with respect to the GR in the light deflection factor, which travels next to an intense gravitational field. This correction leads to a value closer to the measured light deflection by both the eclipse in Sobral in 1919 \cite{dyson,kennefick}, as the more recent measurements \cite{kennefick,fomalont}. They also obtain some results for the weak field approximation, in gravitational waves, and a time dependent cosmological constant, which does not need to be entered as additional term, but being a direct consequence of the non-conservation of the energy-momentum tensor.\par
In this paper, we will explain more clearly some basic concepts of the theory constructed by Resconi et al \cite{resconi1}, through the original inspiration of the gauge theory and the commutator algebra of cavariant derivatives \cite{licata}, for a Riemannian  spacetime, defined through the Levi-Civita connection. We re-obtain the equations of motion of this theory, taking into account all original ingredients used by Resconi, getting a model that possesses an equation of motion different from that presented in \cite{resconi1}. We will analyse in this paper, only the vacuum case in which the metric is static and possesses spherical symmetry. \par
This paper is organized as follows. In the section 2, we present some fundamental concepts for the introduction to the Algebraic Modified Gravity, divided in the subsection 2.1, in which we put out the elements of the GR, and the subsection 2.2, in which we construct the theory. In the section 3, we will present some new solutions for the vacuum case, with a static and spherically symmetric metric, comparing them with the known solutions in the literature and drawing  a preliminary analysis. The conclusion and perspective are presented in the section 4. 


\section{Fundamental Concepts of the Theory}

\subsection{The General Relativity}
Let us firstly rewrite some equations coming from the GR, to be used as the inspiration to the modification made by Resconi et al \cite{resconi1}. The spacetime is considered to have the Riemannian manifold, in which the line element is given by
\begin{equation}
dS^{2}=g_{\mu\nu}dx^{\mu}dx^{\nu}\;,\label{el}
\end{equation}
where $g_{\mu\nu}$ are the components of the metric. The signature of the metric is defined as $(+,-,-,-)$. For making a parallel transport of the vector field $V_{\alpha}(x^{\gamma})$, between two infinitesimally close points, we define the  covariant derivative as being 
\begin{equation}
\nabla _{\mu}V_{\alpha}=\partial_{\mu} V_{\alpha}-\Gamma^{\lambda}_{\;\;\mu\alpha}V_{\lambda}\label{dcov}\;,
\end{equation} 
where $\Gamma^{\lambda}_{\;\;\mu\alpha}$ are the components of the Levi-Civita connection.
If we apply the commutator of the covariant derivatives $\nabla_{\mu}$ and $\nabla_{\nu}$, on a vector $V_{\alpha}$, we get
\begin{equation}
\left[ \nabla_{\mu},\nabla_{\nu}\right]V_{\alpha}=-R^{\lambda}_{\;\;\alpha\mu\nu}V_{\lambda}\;,\label{comutr}
\end{equation}
where we define the Riemann tensor components 
\begin{equation}
R^{\lambda}_{\;\;\alpha\mu\nu}=\partial_{\mu}\Gamma^{\lambda}_{\nu\alpha}-\partial_{\nu}\Gamma^{\lambda}_{\mu\alpha}
+\Gamma^{\sigma}_{\mu\alpha}\Gamma^{\lambda}_{\nu\sigma}-
\Gamma^{\sigma}_{\nu\alpha}\Gamma^{\lambda}_{\mu\sigma}\; .\label{riemann}
\end{equation}
Using two times the commutator of the covariant derivatives $\nabla_{\mu}$, $\nabla_{\alpha}$ and $\nabla_{\beta}$, and making the cyclic subscripts, we get the Jacobi identity 
\begin{equation}
\left[\nabla_{\mu},\left[ \nabla_{\alpha},\nabla_{\beta}\right]\right]+
\left[\nabla_{\alpha},\left[ \nabla_{\beta},\nabla_{\mu}\right]\right]+
\left[\nabla_{\beta},\left[ \nabla_{\mu},\nabla_{\alpha}\right]\right]=0\; ,\label{ijacobi}
\end{equation} 
which in terms of the Riemann tensor in (\ref{comutr}), is written as 
\begin{equation}
\nabla_{\mu}R^{\lambda}_{\;\;\sigma\alpha\beta}+
\nabla_{\alpha}R^{\lambda}_{\;\;\sigma\beta\mu}+
\nabla_{\beta}R^{\lambda}_{\;\;\sigma\mu\alpha}=0\label{ijacobir}\; .
\end{equation}
This latter is called the first Bianchi identity. Contracting $\lambda$ with $\alpha$ in  (\ref{ijacobir}), one obtains 
\begin{equation}
\nabla_{\alpha}R^{\alpha}_{\;\;\sigma\beta\mu}=\nabla_{\beta}R_{\sigma\mu}-\nabla_{\mu}R_{\sigma\beta}\label{1}\; ,
\end{equation}
where $R_{\sigma\mu}$ are the components of the Ricci tensor. Contracting $\sigma$ with  $\beta$, leads to the second Bianchi identity 
\begin{equation}
\nabla_{\alpha}\left[ R^{\alpha}_{\;\;\mu}-\frac{1}{2}\delta^{\alpha}_{\mu}R\right]=0\; ,\label{ibianchi}
\end{equation}
where $R=g^{\mu\nu}R_{\mu\nu}$ is the curvature scalar.
Considering the Einstein equations of the GR
\begin{equation}
R_{\sigma\mu}-\frac{1}{2}g_{\sigma\mu}R=\chi T_{\sigma\mu}\; ,\label{eqE}
\end{equation}
and inserting it in (\ref{1}) one gets
\begin{eqnarray}
\nabla_{\alpha}R^{\alpha}_{\;\;\sigma\beta\mu}=\chi J_{\sigma\beta\mu}\; , \label{qme}
\end{eqnarray}
where
\begin{equation}
J_{\sigma\beta\mu}= \left[ \nabla_{\beta}T_{\sigma\mu}-\nabla_{\mu}T_{\sigma\beta}-\frac{1}{2}\left(g_{\sigma\mu}\nabla_{\beta}T-g_{\sigma\beta}\nabla_{\mu}T\right)\right]\; ,\label{source}
\end{equation}
and  $T$ is the trace of the energy-momentum tensor. The equations (\ref{qme}) are called the old quasi-Maxwellian gravitational equations \cite{nikolai}. Hence, all this is well known and established in the GR. In the next section, we will introduce the fundamental concepts of the theory constructed by Resconi et al \cite{resconi1}. 

\subsection{The Modified Algebraic Gravity}
A modification of the GR made by Camenzind \cite{camenzind} were constructed as an analogy to the the Yang-Mills theory, with the symmetry group $SO(3,1)$. Similarly, inspired by the foundations of the gauge theory, Resconi et al \cite{resconi1} construct a modified theory of gravity, in which the algebra of the covariant derivatives operators adds a term to the equations of motion. In order to present some basic concepts of the theory, we will start applying the commutator of $\nabla_{\mu}$ with the commutator of  $\nabla_{\alpha}$ with  $\nabla_{\beta}$, to a vector field  $K_{\nu}$, we obtain 
\begin{eqnarray}
\left[\nabla_{\mu},\left[\nabla_{\alpha},\nabla_{\beta}\right]\right]K_{\nu} & = & \left(\nabla_{\mu}\left[\nabla_{\alpha},\nabla_{\beta}\right]\right)K_{\nu}-\left[\nabla_{\alpha},\nabla_{\beta}\right]\left(\nabla_{\mu}K_{\nu}\right)\nonumber\\
& = & -\left(\nabla_{\mu}R^{\lambda}_{\;\;\nu\alpha\beta}\right)K_{\lambda}+R^{\lambda}_{\;\;\mu\alpha\beta}\left(\nabla_{\lambda}K_{\nu}\right)\;,
\end{eqnarray}   
and then,
\begin{equation}
\left[\nabla_{\mu},\left[\nabla_{\alpha},\nabla_{\beta}\right]\right]K_{\nu} =  \left(\nabla_{\mu}R^{\lambda}_{\;\;\nu\beta\alpha}\right)K_{\lambda}+R^{\lambda}_{\;\;\mu\alpha\beta}\left(\nabla_{\lambda}K_{\nu}\right)\label{comut}\; .
\end{equation}
This relationship is demonstrated by Resconi et al \cite{resconi2} through the commutative diagrams, as used in category theory. Using the matter source $J_{\sigma\beta\mu}$ in (\ref{source}), we can regain the Einstein equations of GR, for the particular case where the covariant derivative of the vector field $K_{\nu}$ vanishes identically. But firstly, we obtain the general equation of motion for this theory.\par

Considering as matter source, the same of the GR, the $J_{\sigma\beta\mu}$ in (\ref{source}), we can match the expression (\ref{comut}), with the new source condition
\begin{equation}
\left[\nabla_{\mu},\left[\nabla_{\alpha},\nabla_{\beta}\right]\right]K_{\nu} =\chi J_{\mu\alpha\beta}K_{\nu}\label{nsource}\; ,
\end{equation}
which lead to 
\begin{equation}
\chi J_{\mu\alpha\beta}K_{\nu}=\left(\nabla_{\mu}R^{\lambda}_{\;\;\nu\beta\alpha}\right)K_{\lambda}+R^{\lambda}_{\;\;\mu\alpha\beta}\left(\nabla_{\lambda}K_{\nu}\right)\label{2}\; .
\end{equation}
Contracting $\mu$ with $\alpha$, and $\beta$ with  $\nu$, we obtain the following general equation of motion 
\begin{equation}
\nabla_{\mu}\left[ R^{\mu\nu}+\chi \left( T^{\mu\nu}+\frac{1}{2}g^{\mu\nu}T\right)\right]K_{\nu}+R^{\mu\nu}\left(\nabla_{\mu}K_{\nu}\right)=0\; .\label{eq}
\end{equation}
In the particular case where $\nabla_{\mu}K_{\nu}=0$, considering the validity of Einstein's equations (\ref{eqE}), we re-obtain the second Bianchi identity (\ref{ibianchi}) for a non zero vector field $K_{\nu}$. These motion equations are not stemming directly from a Lagrangian or a variational principle, as the case of the Rastall theory \cite{rastall}. All the matter fields, as scalar, vector or spinorial fields, are contained in the energy-momentum tensor,  which gives us the interaction between the matter and the geometry, as in RG.\par
The equation (\ref{eq}) can be interpreted as the equation of motion of a theory which is view as a generalization of the GR, coming from the foundations of gauge theory, in which the added term with respect to the  GR, comes from the covariant derivatives commutator. Indeed, we call this theory of Algebraic Modified Gravity. The vector field $K_{\nu}(x^{\sigma})$ can be interpreted as a field that describes the interaction with a substratum \cite{resconi1}. As in gauge theories, the vector field $K_{\nu}$ can be a gauge field of a certain local (as $SU(2)$ gauge group for exemple) or global (as $SU(3)$) gauge group \cite{moriyasu}. In the case the used , the second term of the sum in (\ref{eq}) is the main coupling term  between the substratum and the gravity, in such a way that   if this term vanishes, the GR is recovered. The crucial point of the difference with the GR here, is the consideration of the source of the Riemann tensor in (\ref{qme}) for the RG,  and being now a source of another operator, like an  eigenvalue equation of the double commutator of covariant derivatives in (\ref{nsource}).


\section{New Solutions in Modified Algebraic Gravity}
To begin with an application of the theory, let us take the case in which the matter content is identically zero, $T^{\mu\nu}=0$ which implies that $J_{\alpha\mu\nu}=0$. The equation (\ref{eq}) for this case is written as
\begin{equation}
\left( \nabla_{\mu}R^{\mu\nu}\right)K_{\nu}+R^{\mu\nu}\left(\nabla_{\mu}K_{\nu}\right)=0\; .\label{eq1}
\end{equation}

As a possible simplification, we can take the eigenvalue condition \cite{resconi1}.
\begin{equation}
\nabla_{\mu}K_{\nu}=\gamma_{\mu}K_{\nu}\label{cond1}\; ,
\end{equation}
which, in (\ref{eq1}) yields 
\begin{equation}
\left(\nabla_{\mu}R^{\mu\nu}+\gamma_{\mu}R^{\mu\nu}\right)K_{\nu}=0\label{3}\;.
\end{equation}
As the vector field is nonzero in general, we have
\begin{equation}
\nabla_{\mu}R^{\mu\nu}+\gamma_{\mu}R^{\mu\nu}=0\label{eq2}\;.
\end{equation}
We can also see that the particular case where the eigenvalues ​​$ \gamma_{\mu}=0$ ($\nabla_{\mu}K_{\nu}=0$), the GR is recovered in the vacuum, which is the second identity Bianchi (\ref{ibianchi}), for $R=$0. For a first approach, we chose a static spacetime with a spherical symmetry. The line element (\ref{el}), of this spacetime can be written as
\begin{equation}
dS^{2}=e^{\nu (r)}dt^{2}-e^{\lambda (r)}dr^{2}-r^{2}\left(d\theta^{2}+\sin^{2}\left(\theta\right)d\phi^{2}\right)\label{ele}\;.
\end{equation}
The components of the inverse metric are given by
\begin{equation}
g^{00}=e^{-\nu (r)}\; , \; g^{11}=-e^{-\lambda (r)}\; , \; g^{22}=-r^{-2}\; , \; g^{33}=-r^{-2}\sin^{-2}\left(\theta\right)\;.\label{mi}
\end{equation}
The nonzero components of the Levi-Civita connection are
\begin{eqnarray}\label{con}
\left\{\begin{array}{lll}\Gamma^{0}_{\;\;10} & = & \frac{1}{2}\nu^{\prime}(r)\; ,\;\Gamma^{1}_{\;\;00}=\frac{1}{2}\nu^{\prime}(r)e^{\nu (r)-\lambda (r)}\; ,\; \Gamma^{1}_{\;\;11}=\frac{1}{2}\lambda^{\prime}(r)\;,\\
\Gamma^{1}_{\;\;22} &=&\Gamma^{1}_{\;\;33}=-re^{-\lambda (r)}\;,\;\Gamma^{2}_{\;\;21} =  \Gamma^{3}_{\;\;31}=r^{-1}\;,\;\Gamma^{2}_{\;\;33}=-\sin\theta\cos\theta\;,\;\Gamma^{3}_{\;\;32}=\cot\theta\;,
\end{array}\right.
\end{eqnarray}
where $^{\prime}$ denotes the derivative with respect to the radial coordinate $r$. The nonzero components of the Ricci tensor are
\begin{eqnarray}\label{ric}
\left\{\begin{array}{lll}R_{00}&=&\frac{1}{4}e^{\nu (r)-\lambda (r)}\left[ 2\nu^{\prime\prime}(r)+\left(\nu^{\prime}(r)\right)^{2}-\nu^{\prime}(r)\lambda^{\prime}(r)+\frac{4}{r}\nu^{\prime}(r)\right]\;,\\
R_{11}&=& -\frac{1}{4}\left[ 2\nu^{\prime\prime}(r)+\left(\nu^{\prime}(r)\right)^{2}-\nu^{\prime}(r)\lambda^{\prime}(r)-\frac{4}{r}\lambda^{\prime}(r)\right]\;,\\
R_{22}&=& \frac{1}{2}e^{-\lambda (r)}\left[ r\lambda^{\prime}(r)+2e^{\lambda (r)}-r\nu^{\prime}(r)-2\right]\;,\\
R_{33}&=&\sin^{2}\theta R_{22}\; .
\end{array}\right.
\end{eqnarray}
We chose a more particular case, where there is only one nonzero  eigenvalue of the equation (\ref{cond1}), $\gamma_{1}(r)$, which depends only on the radial coordinate $r$. We can rewrite the equation (\ref{eq2}) as
\begin{equation}
g^{\mu\sigma}\nabla_{\sigma}R_{\mu\nu}+\gamma^{\mu}R_{\mu\nu}=0\label{coveq2}\;,
\end{equation}
where $\gamma^{\mu}=g^{\mu\sigma}\gamma_{\sigma}$. Using  $\nu=1$, the equation (\ref{coveq2}) becomes 
\begin{eqnarray}
g^{11}\partial_{1}R_{11}&-&R_{11}\Big[g^{00}\Gamma^{1}_{\;\;00}+2g^{11}\Gamma^{1}_{\;\;11}+g^{22}\Gamma^{1}_{\;\;22}+g^{33}\Gamma^{1}_{\;\;33}-\gamma^{1}\Big]\nonumber\\
&-&g^{00}\Gamma^{0}_{\;\;01}R_{00}-g^{22}\Gamma^{2}_{\;\;21}R_{22}-g^{33}\Gamma^{3}_{\;\;31}R_{33}=0\; .\label{eq3}
\end{eqnarray}
Substituting (\ref{mi}), (\ref{con}) and  (\ref{ric}) in  (\ref{eq3}), and after some algebraic manipulation, we obtain the following differential equation
\begin{eqnarray}
&-&8+8e^{\lambda(r)}+2r^{3}\nu^{\prime\prime\prime}(r)-4r\nu^{\prime}(r)+r^{2}\left(\lambda^{\prime}(r)\right)^{2}\left(4+r\nu^{\prime}(r)\right)-4r^{2}\lambda^{\prime\prime}(r)\nonumber\\
&-&r^{3}\nu^{\prime}(r)\lambda^{\prime\prime}(r)+4r^{2}\nu^{\prime\prime}(r)+2r^{3}\nu^{\prime}(r)\nu^{\prime\prime}(r)
-r^{2}\lambda^{\prime}\left[4\nu^{\prime}(r)+r\left(\nu^{\prime}(r)\right)^{2}+3r\nu^{\prime\prime}(r)\right]\nonumber\\
&+&e^{\lambda(r)}r^{2}\gamma^{1}(r)\Big[\lambda^{\prime}(r)\left(4+r\nu^{\prime}(r)\right)-r\left(\nu^{\prime}(r)\right)^{2}+2\nu^{\prime\prime}(r) \Big]=0\; .\label{eq4}
\end{eqnarray}
This is a non-linear differential equation of third order, but it can be simplified to obtain exact solutions, as we shall see later. Choosing the system of quasi-global coordinates where the metric functions satisfy $\nu(r)=-\lambda(r)$($g_{11}=-g^{00}$), and making the redefinition
\begin{equation}
\nu(r)=-\lambda(r)=\ln\left[1+f(r)\right]\;,\label{fr}
\end{equation}
the differential equation (\ref{eq4}), after some algebraic manipulations, becomes
\begin{eqnarray}
r^{3}f^{\prime\prime\prime}(r)-4f(r)-2rf^{\prime}(r)\left[1+\frac{r\gamma^{1}(r)}{1+f(r)}\right]+
r^{2}f^{\prime\prime}(r)\left[4-\frac{r\gamma^{1}(r)}{1+f(r)}\right]=0\label{eq5}\;.
\end{eqnarray}
Remembering that $\gamma^{1}(r)=g^{11}\gamma_{1}(r)=-\left(1+f(r)\right)\gamma_{1}(r)$, the equation (\ref{eq5})  leads to 
\begin{eqnarray}
r^{3}f^{\prime\prime\prime}(r)-4f(r)-2rf^{\prime}(r)\left[1-r\gamma_{1}(r)\right]+
r^{2}f^{\prime\prime}(r)\left[4+r\gamma_{1}(r)\right]=0\label{eq6}\;.
\end{eqnarray}
This differential equation is now linear and of third order. For each type of $\gamma_{1}(r)$, we get a possible solution of this equation. We will show here some interesting cases:
\begin{enumerate}
\item  The first model we will take is that of GR, ie, $\gamma_{1}(r)=0$. The equation (\ref{eq6}) becomes 
\begin{eqnarray}
r^{3}f^{\prime\prime\prime}(r)-4f(r)-2rf^{\prime}(r)+
4r^{2}f^{\prime\prime}(r)=0\label{eq7}\;.
\end{eqnarray}
The general solution for this equation is given by
\begin{equation}
f(r)=\frac{c_{1}}{r}+\frac{c_{2}}{r^{2}}+c_{3}r^{2}\;, \label{sol1}
\end{equation}
where $c_{1}$, $c_{2}$ and $c_{3}$ are real constants. When we choose $c_{1}=-2M$, $c_{2}=q^{2}$ and $c_{3}=\Lambda/3$, with $M$ being the total mass, $q$ the charge and $\Lambda$ the cosmological constant, we have the solution of Reissner-Nordstrom-(anti) de Sitter (RN-(A)dS). The equation (\ref{eq7}) is obtained directly from the second Bianchi identity (\ref{ibianchi}) for the vacuum ($R=0$). Hence, (\ref{sol1}) is solution of the second Bianchi identity (\ref{ibianchi}) for the vacuum, ie, a particular solution of GR.


\item Now let us take the model where $\gamma_{1}(r)=-A$, with $A$ is a real constant. The equation (\ref{eq6}) becomes
\begin{equation}
r^{3}f^{\prime\prime\prime}(r)-4f(r)+r^{2}f^{\prime\prime}(r)\left(4-Ar\right)-2rf^{\prime}(r)\left(1+Ar\right)=0\label{eq8}\; .
\end{equation}
A particular solution of this equation in order to eliminate uninteresting terms, is given by
\begin{equation}
f(r)=\frac{c_{2}}{A^{2}}+\frac{c_{1}}{r}-\frac{6c_{2}}{A^{4}r^{2}}+\frac{4c_{2}}{A^{3}r}\ln(r)\;, \label{sol2}
\end{equation}
where $c_{1}$ and $c_{2}$ are real constants. This solution is not asymptotically Minkowskian for $c_{2}\neq 0$. When $c_{1}=-2M$ and $c_{2}=-q^{2} A^{4}/6$, the solution resembles to that of  RN with a complicated logarithmic term. This term appears in the solutions of $(1+2)$-Dimensional Teleparallel gravity \cite{kawai} and Symmetric Teleparallel Gravity \cite{adak}. Note also that This sort of term has been obtained, in $f(R)$ theory, in the limit of the fourth post-Newtonian order correction \cite{mariafelicia} (Eq. $(16.100)$). The signature of the metric in the asymptotic limit depends on the constants $c_{2}$ and $A$. When $c_{2}=-q^{2}A^{4}/6 \;$ and $\;-1<(\left|q\right|A/\sqrt{6})<1$, the  asymptotic signature is  $(+,-,-,-)$, and otherwise $(-,+,-,-)$.
\item When $\gamma_{1}(r)=-Ar$, with $A$ being a real constant, the equation (\ref{eq6}) becomes 
\begin{equation}
r^{3}f^{\prime\prime\prime}(r)-4f(r)+r^{2}f^{\prime\prime}(r)\left(4-Ar^{2}\right)-2rf^{\prime}(r)\left(1+Ar^{2}\right)=0\label{eq9}\; .
\end{equation}
A particular solution of (\ref{eq9}) is given by 
\begin{equation}
f(r)=-\frac{c_{2}}{A}+\frac{c_{1}\sqrt{2A}}{r}+\frac{2c_{2}}{A^{2}r^{2}}\;, \label{sol3}
\end{equation}
where $c_{1}$ and $c_{2}$ are real constants. In the same way as for the previous solution, this solution is not asymptotically Minkowskian for $c_{2}\neq 0 $. When $c_{1}=-2M/\sqrt{2a}\,$ and $c_{2}=(qA)^{2}/2$, the solution is similar to that of RN. The signature of the metric in the asymptotic limit depends on the constant $c_{2}$ and $A$. When $c_{2}=(qA)^{2}/2\;$ and $\;-1 <(\left|q\right|A/\sqrt{2})<1$, the asymptotic signature is $(+,-,-,-)$, otherwise is $(-,+,-,-)$.
\item When $\gamma_{1}(r)=-A/r$, with $A$ being a real constant,     the equation (\ref{eq6}) becomes 
\begin{equation}
r^{3}f^{\prime\prime\prime}(r)-4f(r)+r^{2}f^{\prime\prime}(r)\left(4-A\right)-2rf^{\prime}(r)\left(1+A\right)=0\label{eq10}\; .
\end{equation}
The general solution of (\ref{eq10}) is given by 
\begin{equation}
f(r)=\frac{c_{1}}{r}+c_{2}r^{\frac{1}{2}\left(A-\sqrt{A^{2}+16}\right)}+c_{3}r^{\frac{1}{2}\left(A+\sqrt{A^{2}+16}\right)}\;, \label{sol4}
\end{equation}
where $c_{1}$, $c_{2}$ and $c_{3}$ are real constants. This solution is a generalization of the solution of RN-(A)dS. When $A=0$ ($\gamma_{1}(r)=0$), the GR is recovered, with the solution of RN-(A)dS in (\ref{sol1}). The exponent $r$ of the constants terms $c_{2}$ and $c_{3}$ can take real values ​​in general. Similar solutions arise in modified gravity, as in the case of $f(R)$ theory with $R$ being the scalar curvature. The powers of the radial coordinate $r$ in the Clifton paper \cite{clifton}, depends on the value of $\delta$, which is power of $R$ ($f(R)=R^{1+\delta}$). Note that the solution proposed here is more general in the sense that power of $r$ is general and also there exists a mass term of order $1/r$.
\item When  $\gamma_{1}(r)=-A/r^{2}$, with $A$ a real constant, the equation (\ref{eq6}) reads
\begin{equation}
r^{3}f^{\prime\prime\prime}(r)-4f(r)+rf^{\prime\prime}(r)\left(4r-A\right)-2f^{\prime}(r)\left(r+A\right)=0\label{eq11}\; .
\end{equation}
The general solution of (\ref{eq11}) is given by 
\begin{eqnarray}
f(r)&= &\frac{A^{2}c_{2}}{A^{2}-4A+6}+\frac{1}{r}\left[c_{1}-\frac{A^{2}-2A+2}{A^{2}-4A+6}c_{2}\right]-\frac{2Ac_{2}}{A^{2}-4A+6}r\nonumber\\
&&+\frac{2c_{2}}{A^{2}-4A+6}r^{2}+\frac{c_{3}r^{2}}{A^{4}(A^{2}-4A+6)}e^{-A/r}\;, \label{sol5}
\end{eqnarray}
where  $c_{1}$, $c_{2}$ e $c_{3}$ are real constants. As in the two previous solutions, this solution is not asymptotically Minkowskian for $c_{2}\neq 0$ and $c_{3}\neq 0$. When $c_{1}=-2M+\frac{\Lambda}{6}\left(A^{2}-2A+2\right)$ and $c_{2}=\frac{\Lambda}{6}\left(A^{2}-2A+2\right)$, the solution is similar with that of S-(A)dS, with a linear term and other exponential one that differentiates these solutions. An interesting term is the linear one, in (\ref{sol5}), which is connected with the Rindler acceleration, which is more appreciable for large distances \cite{daniel}. This Rindler acceleration $a$ with $a=\frac{-Ac_{2}}{A^{2}-4A+6}$, is in the  direction of the source when $a>0$ and forth from the source when $a<0$. In fact, this term appears firstly in the solution of a theory whose Lagrangian has the quadratic contraction of the Weyl tensor, or a quadratic term of the scalar curvature $R$, added to a quadratic contraction term of the  Ricci tensor. This solution was first obtained by Kazanas and Mannheim \cite{kazanas}, where, according to their parameters, $\gamma=-Ak$, $\beta=-\frac{A}{6}$, $k=\frac{2c_{2}}{A^{2}-4A+6}$ and $c_{3}=0$. They showed that in this case, the solution interpolates the Schwarzschild's one and also that of Robertson-Walker. This is not view directly as in \cite{kazanas} with $\beta=0$, which in our case leads to $A=0$. In our case, we must first assume that the constant term in (\ref{sol5}) is much smaller than unity, which is acceptable due to the fact that it is proportional to the cosmological constant $\Lambda$ in a particular case. Then, we choose $c_{1}=\frac{\Lambda}{6}\left(A^{2}-2A+2\right)$, such that the mass is zero, and then can be compared with the case of \cite{kazanas}. Just as in \cite{daniel}, the linear term in $r$ starts to dominate for large distances. Note that our solution is more general because the exponential term. This linear term also has been obtained for the second order post-Newtonian corrections in $f(R)$ theory \cite{mariafelicia} (Eq $(16.87)$). Also, note that the exponential term, of the form of Yukawa correction, has been found from the second order post-Newtonian corrections in the $f(R)$ theory \cite{mariafelicia} (Eq. $(16.74)$). However, it is important to mention that we have here the inverse of $r$ in the exponential argument, and then the correction would be observed in a cosmological constant term, proportional to $r^2$, which is not the case of the $f(R)$ theory where the correction appears in a mass term (proportional to $r^{-1}$). 
\item We could have several cases for the function $\gamma_{1}(r)=-Ar^{n}$, with an integer $n$, but all provide us very long solutions of little interest, which are not necessary to be shown here. We only put out these possibilities. A simple example is when $n=-3$, ie $\gamma_{1}(r)=-A/r^{3}$. The general solution of (\ref{eq6}) is
\begin{eqnarray}
f(r) &=& -\frac{Ac_{2}}{A-2}+\frac{1}{r}\left[c_{1}-\frac{2-3A}{3(A-2)}c_{2}\right]+\frac{2c_{2}}{3(A-2)}r^{2}\nonumber\\
&&-\frac{c_{3}}{6(A-2)A^{2}r}\left[2r\left(r^{2}-A\right)e^{-\frac{A}{2r^{2}}}-A^{3/2}\sqrt{2\pi} Er\left(\sqrt{\frac{A}{2}}\frac{1}{r}\right)\right]\label{sol6}\; ,
\end{eqnarray}     
where $Er(x)=\frac{2}{\sqrt{\pi}}\int e^{-x^{2}}dx$. Using the particular case where $c_{3}=0$,this solution is similar to that of S-(A)dS, for $c_{1}=-2M+\frac{\Lambda}{6}(2-3A)$ and  $c_{2}=\frac{\Lambda}{2}(A-2)$. 

\end{enumerate}

\section{Conclusion}

We present some basic concepts of the theory called Modified Algebraic Gravity through known elements in the GR and taking as inspiration the foundations of gauge theory. Our goal in revisiting this theory is to investigate the reconstruction of some black holes solutions, well known and already obtained from the GR and the most usual modified theories of gravitation, as f(R) and f(T). We obtained the equations of motion for this theory and chose as a first application the static vacuum with spherical symmetry. The equations for this particular case are non linear differential equations of third order, coming from the Bianchi identities coupled with a term proportional to the Ricci tensor. Considering the covariant derivative of the gauge field $ K_{\nu}$, which represents the interaction of gravity with the substratum, providing an equation as that of eigenvalue, we might rewrite the equation of motion as a linear differential equation of third order. In such a way, we shown some consistency with  the well established results,  and then getting a trustable  direct analogy with the usual modified theories, allowing us to obtain new results.
\par
We divided the solutions obtained in this paper into six cases. The first, is when the eigenvalue $\gamma_{1}(r)=0$, and we re-obtained the solution of RN-(A)dS of the GR. The second is when $\gamma_{1}(r)=-A$, with $A\in\mathcal{R}$, we get a solution that generalizes the RN by a logarithm term, and the possibilities of asymptotic signature, $r\rightarrow+\infty$, and $(+,-,-,-)$\,and\, $(-,+,-,-)$. The third, when $\gamma_{1}(r)=-Ar$, which results in a solution that resembles to that of RN, with the possibility of asymptotic signature  $(+,-,-,-)$ and $ (- ,+,-,-)$. The fourth case is when $\gamma_{1}(r)=-A/r $, which results in a solution that generalizes the RN-(A)dS in any real power of radial coordinate $r$. This solution generalizes the Clifton \cite{clifton} by a mass term proportional to $1/r$.
The fifth case is when $\gamma_{1}(r)=-A/r^{2}$, which results into a solution that generalizes the S-(A) dS for two terms, one linear in $r$, which represents  the Rindler acceleration, as in \cite{daniel}, and another which present an exponential form. The solution, in a particular case, can interpolate the Schwarzschild and Robertson-Walker ones, as shown in \cite{kazanas}. We still have several possibilities for solutions for $\gamma_{1}(r)=-Ar^{n}$, with $n$ an integer, but which are very long and are not explicitly presented here. We  only put out the case where $n=-3$, the sixth case, as the simplest example, resulting into a solution similar to that of S-(A) dS, for $c_{3}=0$ in (\ref{sol6}). \par 
Then, we wrap up that with this theory, we were able to reproduce partially and even fully, at least in the vacuum, some results previously obtained by the above usual modified theories of gravitation. In such a way, we believe that with this theory, further investigations could lead to more information which are not clear until now about the dark sector of the universe. Note that this kind of work, in investigating the dark sector of the universe, has previously been made by Resconi \cite{licata}. However, there are still many things to do in this way. Also, there is still a wide range of possible solutions, both static and cosmological to this theory. These aspects will be undertaken in our future works, as well as the introduction of a more complex structure for the field gauge $K_{\nu}$ in (\ref{eq}), as $U(1)$ or $SU(2)$ gauge groups, for testing with vehemence this theory and trying to make some explanations to some points that continue being a challenge until now about the dark sector of the universe, and also obtaining new results in local gravitation and astrophysics, consistent with the observational data. 

\vspace{0,5cm}

{\bf Acknowledgement:} M. H. Daouda thanks CNPq/TWAS for financial support. M. E. Rodrigues
thanks UFES for the hospitality during the development of this work. M. J. S. Houndjo thanks CNPq
for partial financial support. 


\end{document}